\documentclass{PoS}

\title{Lattice simulation of\\ultracold atomic Bose-Fermi mixtures}

\ShortTitle{Lattice simulation of ultracold atomic Bose-Fermi mixtures}

\author{Arata~Yamamoto\\
Theoretical Research Division, Nishina Center, RIKEN, Saitama 351-0198, Japan\\
E-mail: \email{arayamamoto@riken.jp}}

\abstract{
Bose-Fermi mixtures have been recently realized and invesitigated in ultracold atomic experiments.
We formulate quantum Monte Carlo simulation of Bose-Fermi mixtures on the (3+1)-dimensional lattice.
As its first application, we analyze the boson-fermion pair correlation and the phase diagram of the Bose-Einstein condensation.
}

\FullConference{The XXX International Symposium on Lattice Field Theory\\
June 24-29, 2012\\
Cairns Convention Centre, Cairns, Australia}

\begin{document}

\section{Introduction}
One frontier of ultracold atomic physics is multi-component quantum system, such as Bose-Fermi mixtures.
The most famous example is a $^4$He-$^3$He mixture.
Recently, many kinds of Bose-Fermi mixtures have been investigated in ultracold atomic experiments.
In particular, Bose-Fermi mixtures can be trapped on a three-dimensional optical lattice.
The mixtures of $^{87}$Rb-$^{40}$K \cite{Gunter:2006,Ospelkaus:2006,Best:2009zz}, $^{170}$Yb-$^{173}$Yb and $^{174}$Yb-$^{173}$Yb \cite{Sugawa:2011} were examined on the optical lattice.
Since the optical lattice is a real physical lattice, we can exactly study lattice physics in laboratory.

On the theoretical side, we can exactly reproduce the lattice physics in quantum Monte Carlo simulations.
Bose-Fermi mixtures have been studied in quantum Monte Carlo simulations in 1+1 dimensions \cite{Takeuchi:2006,Pollet:2006,Sengupta:2007,Hebert:2007,Pollet:2008,Varney:2008,Masaki:2008}.
All the previous simulations were done in the world-line formalism \cite{Hirsch:1983wk} or its extensions \cite{Prokofev:1998,Sandvik:1999}.
The world-line formalism is an exact scheme in 1+1 dimensions.
In higher dimensions, however, there is the sign problem which originates from antisymmetric property of fermions.
For this reason, there was no quantum Monte Carlo study of Bose-Fermi mixtures in 3+1 dimensions.

In this study, we perform quantum Monte Carlo simulation of Bose-Fermi mixtures on the (3+1)-dimensional lattice \cite{Yamamoto:2012wy}.
We adopt the same framework as the lattice QCD simulation.
In this framework, there is no sign problem on fermions.
We use the lattice unit and drop the lattice constant throughout this paper.

\section{Formalism}
We consider one-component boson field $\Phi(\vec{x},\tau)$ and two-component fermion field $\Psi_\uparrow(\vec{x},\tau)$ and $\Psi_\downarrow(\vec{x},\tau)$.
The formalism is based on the path integral in terms of the Euclidean action.
The generating functional is
\begin{eqnarray}
Z &=& \int D\Phi^* D\Phi D\Psi_\uparrow^* D\Psi_\uparrow D\Psi_\downarrow^* D\Psi_\downarrow \ e^{-S} \nonumber \\
&=&  \int D\Phi^* D\Phi \ \det K_\uparrow \det K_\downarrow e^{-S_B} \ ,
\end{eqnarray}
This system is similar to the two-flavor QCD, in which the boson field is gluon and the fermion fields are u-quark and d-quark.
We adopted the hybrid Monte Carlo algorithm, which is frequently used in the lattice QCD simulation.

For the lattice action, we consider the Bose-Fermi Hubbard model.
The naive form of the action is
\begin{eqnarray}
S &=& S_B +S_F +S_{BF} \\
S_B &=& \sum_{\vec{x},\tau} \Big[ \Phi^*(\vec{x},\tau) \{ \Phi(\vec{x},\tau) - \Phi(\vec{x},\tau-1) \} -\mu_B n_B(\vec{x},\tau) \nonumber \\
&& - t_B \sum_{j=1}^3 \{ \Phi^*(\vec{x},\tau) \Phi(\vec{x}+\vec{e}_j,\tau)+ \Phi^*(\vec{x},\tau) \Phi(\vec{x}-\vec{e}_j,\tau) \} + U_B n_B(\vec{x},\tau) \{n_B(\vec{x},\tau) -1\} \Big] \\
S_F &=& \sum_{\vec{x},\tau,\sigma} \Big[ \Psi_\sigma^*(\vec{x},\tau) \{ \Psi_\sigma(\vec{x},\tau) - \Psi_\sigma(\vec{x},\tau-1) \} - \mu_{F\sigma} n_{F\sigma}(\vec{x},\tau) \nonumber \\
&& - t_{F\sigma} \sum_{j=1}^3 \{ \Psi_\sigma^*(\vec{x},\tau) \Psi_\sigma(\vec{x}+\vec{e}_j,\tau) + \Psi_\sigma^*(\vec{x},\tau) \Psi_\sigma(\vec{x}-\vec{e}_j,\tau) \} \Big] \\
S_{BF} &=& \sum_{\vec{x},\tau,\sigma} U_{BF} n_B(\vec{x},\tau) n_{F\sigma}(\vec{x},\tau)
\end{eqnarray}
with $n_B(\vec{x},\tau) =  \Phi^*(\vec{x},\tau) \Phi(\vec{x},\tau)$ and $n_{F\sigma}(\vec{x},\tau) =  \Psi_\sigma^*(\vec{x},\tau) \Psi_\sigma(\vec{x},\tau)$.

We list several remarks on this action:
\begin{itemize}

\item This action has the sign problem on the imaginary-time derivative term of the boson field.
This is understood in the Fourier transformation as
\begin{eqnarray}
\sum_{\vec{x},\tau} \Phi^*(\vec{x},\tau) \{ \Phi(\vec{x},\tau) - \Phi(\vec{x},\tau-1) \}
= N_\tau \sum_{\vec{x},k} \{ 1-e^{-i\omega_{Bk}} \} \tilde{\Phi}^*(\vec{x},k) \tilde{\Phi}(\vec{x},k) \ .
\end{eqnarray}
The boson Matsubara frequency is $\omega_{Bk} = 2k\pi T$.
To avoid this sign problem, we adopted {\it the zero-frequency approximation}.
In this approximation, nonzero-frequency modes $\tilde{\Phi}(\vec{x},k\ne 0)$ are set to zero, and the boson field is independent of imaginary time as $\Phi(\vec{x},\tau) = \tilde{\Phi}(\vec{x},k=0) \equiv \Phi(\vec{x})$.
This approximation is a priori justified in high-temperature limit or near the critical temperature.
In general case, we can check the validity of the approximation by adding a few lowest frequency modes, e.g., in the reweighting method.
This sign problem is characteristic in the non-relativistic theory.
There is no sign problem in the relativistic scalar theory because the imaginary-time derivative term is bilinear.

\item We do not consider the fermion self-interaction term for simplicity.
It is straightforward to treat the fermion self-interaction term by the Hubbard-Stratonovich transformation and the auxiliary field.

\item We take the same hopping parameter and the same chemical potential for the two fermions, $t_F\equiv t_{F\uparrow} =t_{F\downarrow}$ and $\mu_F\equiv \mu_{F\uparrow} =\mu_{F\downarrow}$, because the hybrid Monte Carlo algorithm becomes simple.
The two fermions are degenerated.

\item The imaginary-time derivative is discretized to the backward difference.
The chemical potential and the interaction are multiplied to the backward hopping term, according to Refs.~\cite{Hasenfratz:1983ba,Chen:2003vy}.
For example,
\begin{eqnarray}
&&\Psi^*(\vec{x},\tau) [ \Psi(\vec{x},\tau) - \Psi(\vec{x},\tau-1) ] + \{ -\mu_F + U_{BF} n_B(\vec{x}) \} \Psi^*(\vec{x},\tau) \Psi(\vec{x},\tau) \nonumber \\
&&\to \Psi^*(\vec{x},\tau) [ \Psi(\vec{x},\tau) + \{ -e^{\mu_F} + U_{BF} n_B(\vec{x}) \} \Psi(\vec{x},\tau-1) ] \ .
\end{eqnarray}

\item The most time-consuming part of the simulation is the inversion of the fermion matrix.
We used the BiCGstab solver for the matrix inversion.
We performed the Fourier transformation to the frequency space for preconditioning, i.e., for accelerating the convergence of the solver.
After the Fourier transformation, the fermion matrix becomes diagonal in the frequency space.
Because the temporal lattice size $N_\tau$ is much larger than the spatial lattice size $N_s$, this preconditioning greatly improves the convergence of the solver.

\end{itemize}

From the above consideration, the final form of the lattice action is
\begin{eqnarray}
S_B &=& N_\tau \sum_{\vec{x}} \Big[ (1-e^{\mu_B}) n_B(\vec{x}) \nonumber \\
&& - t_B \sum_{j=1}^3 \{ \Phi^*(\vec{x}) \Phi(\vec{x}+\vec{e}_j)+ \Phi^*(\vec{x}) \Phi(\vec{x}-\vec{e}_j) \} + U_B n_B(\vec{x}) \{n_B(\vec{x}) -1\} \Big] \\
S_F &=& N_\tau \sum_{\vec{x},k,\sigma} \Big[ ( 1-e^{\mu_F}e^{-i\omega_{Fk}} ) \tilde{n}_{F\sigma}(\vec{x},k) \nonumber \\
&& - t_F \sum_{j=1}^3 \{ \tilde{\Psi}_\sigma^*(\vec{x},k) \tilde{\Psi}_\sigma(\vec{x}+\vec{e}_j,k) + \tilde{\Psi}_\sigma^*(\vec{x},k) \tilde{\Psi}_\sigma(\vec{x}-\vec{e}_j,k) \} \Big]  \\
S_{BF} &=& N_\tau \sum_{\vec{x},k,\sigma} U_{BF} e^{-i\omega_{Fk}} n_B(\vec{x}) \tilde{n}_{F\sigma}(\vec{x},k) \ .
\end{eqnarray}
The fermion Matsubara frequency is $\omega_{Fk} = (2k-1)\pi T$.

We can prove the positivity and the reality of the fermion determinant.
In the fermion matrix, the complex factor is only $e^{-i\omega_{Fk}}$ and other parts give real eigenvalues.
Therefore, the fermion determinant is
\begin{eqnarray}
\det K_\sigma = \prod_k (A+ e^{-i\omega_{Fk}}B) = \prod_{\sin \omega_{Fk}>0} [ (A+ B\cos \omega_{Fk})^2 + B^2 \sin^2 \omega_{Fk} ] > 0 \ ,
\end{eqnarray}
where $A$ and $B$ are real numbers.

There are many simulation parameters in this system.
We fixed the hopping parameters $t_B = t_F = 0.01$, the chemical potentials $\mu_B = \mu_F =0$, the boson self-interaction $U_B = 0.1$, and the spatial lattice volume $N_s^3 = 10^3$.
We varied the boson-fermion interaction $U_{BF} = -0.1$ to 0.1 and temperature $T=1/N_\tau= 0.01$ to 0.05.

\section{Result}
\begin{figure}[b]
\begin{center}
\includegraphics[scale=0.8]{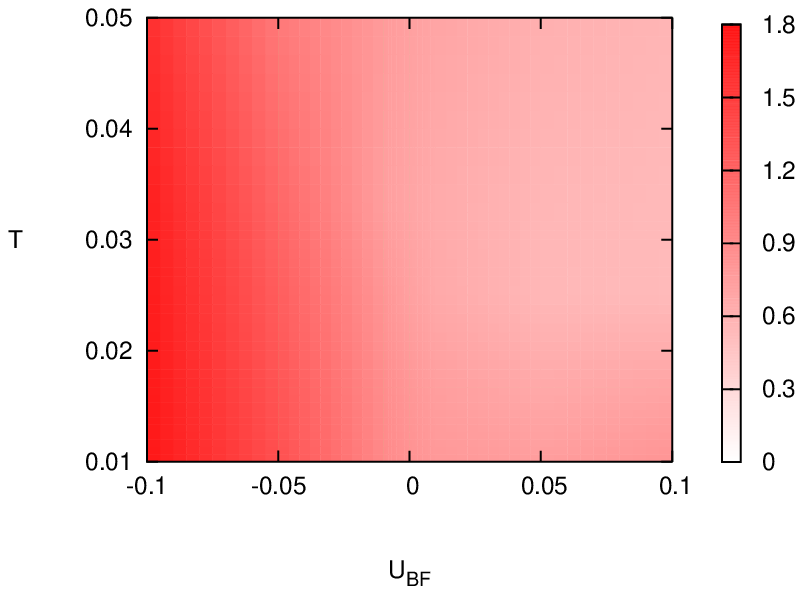}
\includegraphics[scale=0.8]{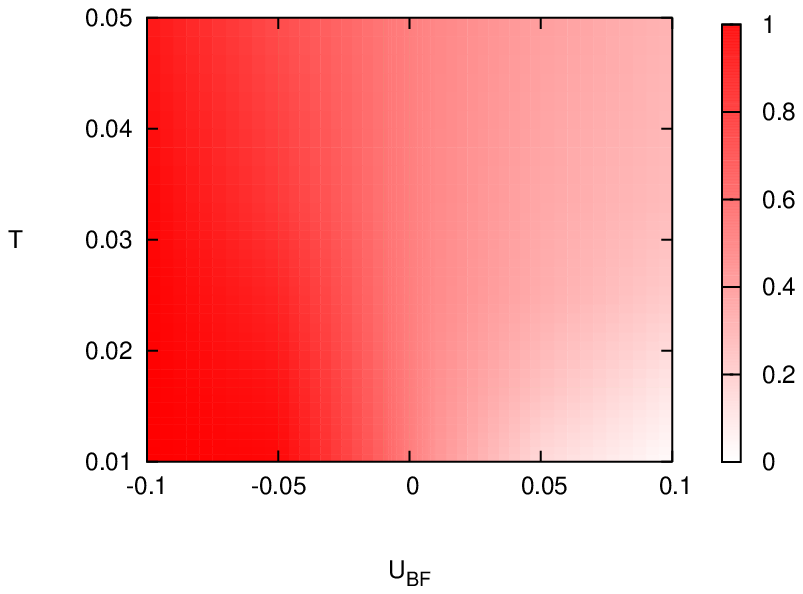}
\caption{\label{fig1}
Left: the boson number density $\langle n_B \rangle$.
Right: the fermion number density $\langle n_{F} \rangle$.
}
\end{center}
\end{figure}

Because the simulation is in the grand-canonical formalism, number densities are not fixed.
The number densities are functions of chemical potentials and other parameters.
In Fig.~\ref{fig1}, we show the boson number density $\langle n_B \rangle$ and the fermion number density $\langle n_{F} \rangle$.
The number densities depend on the boson-fermion interaction $U_{BF}$.
The number densities are decreasing functions of $U_{BF}$ because the chemical potentials are replaced as $\mu_B \to \mu_B - U_{BF} n_F$ and $\mu_F \to \mu_F - U_{BF} n_B$ at tree level.
Since the free fermion has the particle-hole symmetry, the fermion is half-filling, i.e., $\langle n_{F} \rangle = 0.5$ at $U_{BF}=0$.
On the other hand, the boson does not have the particle-hole symmetry at finite $U_B$.
Thus, the regions of $U_{BF}>0$ and $U_{BF}<0$ are not symmetric.

\begin{figure}[t]
\begin{center}
\includegraphics[scale=0.95]{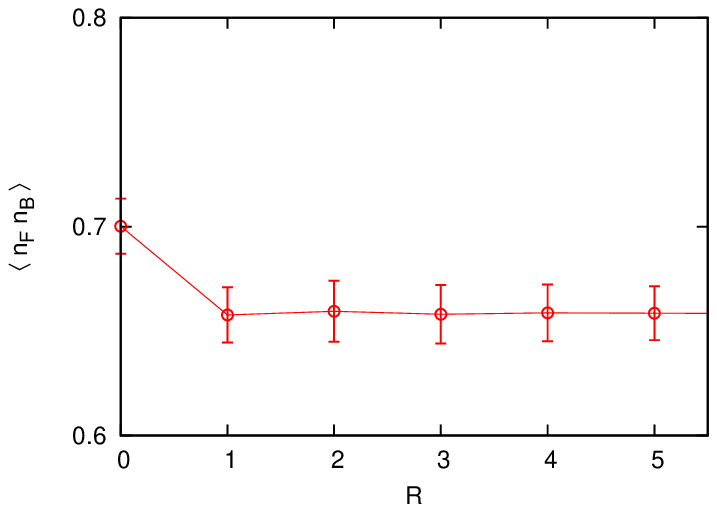}
\includegraphics[scale=0.95]{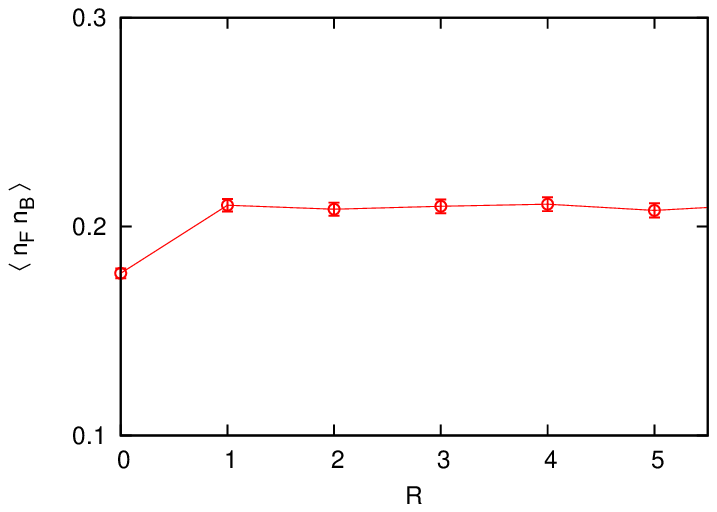}
\caption{\label{fig2}
The boson-fermion correlation function $\langle n_B (\vec{x}) n_F (\vec{y}) \rangle$ as a function of the distance $R= |\vec{x}-\vec{y}|$.
Left: the attractive boson-fermion interaction $U_{BF}=-0.05$.
Right: the repulsive boson-fermion interaction $U_{BF}=0.05$.
}
\end{center}
\end{figure}

In this setup, we measured the boson-fermion pair correlation $\langle n_B (\vec{x}) n_F (\vec{y}) \rangle$.
At $\vec{x} = \vec{y}$, this quantity is called the pair occupancy.
The physical interpretation of the pair occupancy is a probability to find boson-fermion pairs in a single lattice site.
In Fig.~\ref{fig2}, we plot the boson-fermion pair correlation as a function of the distance $R= |\vec{x}-\vec{y}|$.
The temperature is $T=0.05$.
In the attractive case $U_{BF}<0$, the pair correlation is enhanced at $R=0$, and thus the formation of boson-fermion pairs is favored.
In the repulsive case $U_{BF}>0$, the pair correlation is reduced at $R=0$, and thus the bosons and the fermions tend to separate.

\begin{figure}[t]
\begin{center}
\includegraphics[scale=0.95]{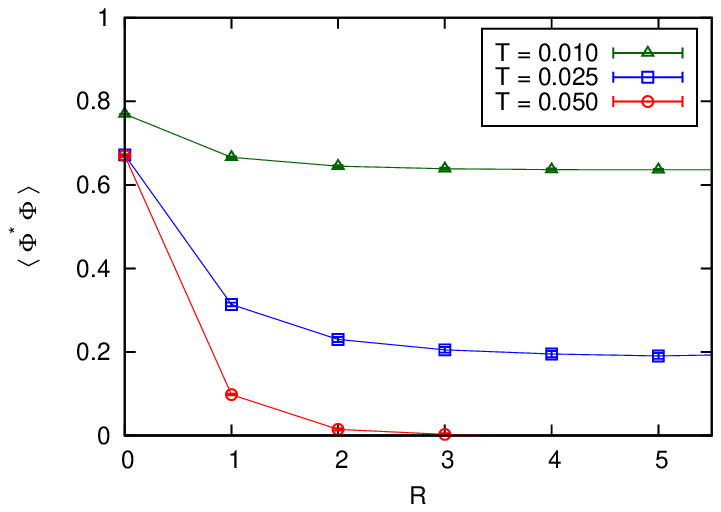}
\includegraphics[scale=0.8]{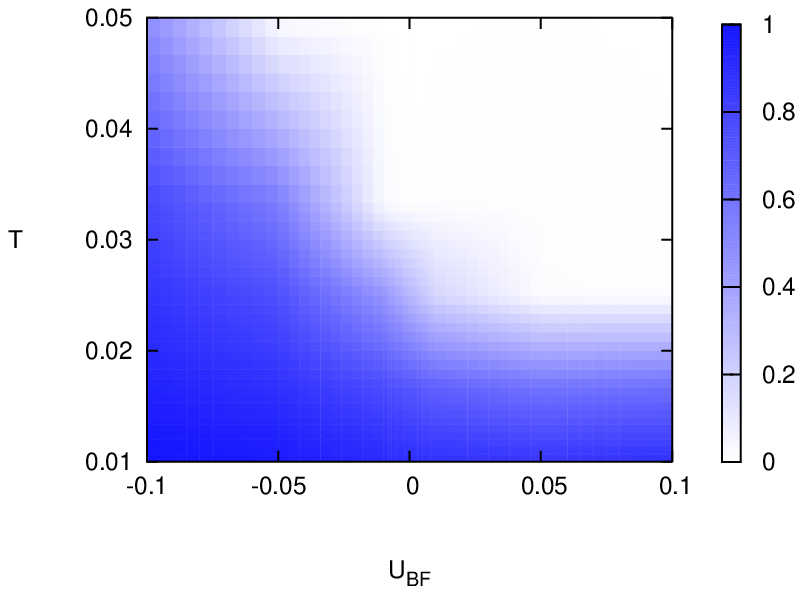}
\caption{\label{fig3}
Left: the boson propagator $\langle \Phi^* (\vec{x}) \Phi (\vec{y}) \rangle$ as a function of the distance $R= |\vec{x}-\vec{y}|$ at $U_{BF}=0$.
Right: the phase diagram of the Bose-Einstein condensation.
}
\end{center}
\end{figure}

Next we analyze the Bose-Einstein condensation.
The Bose-Einstein condensation is identified from the long-range behavior of the boson propagator $\langle \Phi^* (\vec{x}) \Phi (\vec{y}) \rangle$, which is so-called the off-diagonal long-range order.
As shown in the left panel of Fig.~\ref{fig3}, the boson propagator drops to zero at high temperature.
At low temperature, zero-momentum mode of the boson field appears and the boson propagator has a nonzero expectation value in $R \to \infty$.
This expectation value corresponds to the Bose-Einstein condensation density.
In the right panel of Fig.~\ref{fig3}, we draw the condensate fraction
\begin{eqnarray}
\frac{\langle n_{B0} \rangle}{\langle n_B \rangle} = \frac{\langle \Phi^* (\vec{x}) \Phi (\vec{y}) \rangle_{R=N_s/2}}{\langle \Phi^* (\vec{x}) \Phi (\vec{y}) \rangle_{R=0}} \ .
\end{eqnarray}
Despite the fixed boson self-interaction $U_B$, the phase transition temperature is changed by the boson-fermion interaction $U_{BF}$.
We see that the fermion-induced interaction affects the Bose-Einstein condensation.
This is similar to the confinement-deconfinement phase transition in QCD.
Color confinement is a gluon phenomenon, but its phase transition temperature is changed by the dynamical quark effect.
The current simulation was performed in a fixed spatial volume.
For precisely determining the phase boundary, we need to perform the finite size scaling.

\section{Summary}

We have formulated the quantum Monte Carlo simulation of Bose-Fermi mixtures on the (3+1)-dimensional lattice.
We have calculated the boson-fermion pair correlation and the phase diagram of the Bose-Einstein condensation.
Physics in 3+1 dimensions differs from physics in 1+1 dimensions.
For example, the Bose-Einstein condensation cannot be observed in 1+1 dimensions.
We should note that these are experimental observables in ultracold atomic physics.
The pair occupancy was experimentally measured on an optical lattice \cite{Sugawa:2011}.
The fermion-induced effect on the Bose-Einstein condensation was observed in the interference pattern on an optical lattice \cite{Gunter:2006}.

\section*{Acknowledgments}
The author is grateful to T.~Abe, T.~Hatsuda, and R.~Seki for useful discussions.
The author is supported by the Special Postdoctoral Research Program of RIKEN.
The lattice simulations were carried out on NEC SX-8R in Osaka University.

\end{document}